## Personal applications, based on moveable / resizable elements

All the modern day applications have the interface, absolutely defined by the developers.  The use of adaptive interface or dynamic layout allows some variations, but even all of them are predetermined on the design stage, because the best reaction (from designer's view) on any possible users' movement was hardcoded.  But there is a different world of applications, totally constructed on moveable / resizable elements; such applications turn the full control to the users.  The crucial thing in such programs is that not something but everything must become moveable and resizable.  This article describes the features of such applications and the algorithm behind their design.

Programs are developed by one or several people and may have from several users to several millions.  For all the commercial products, the number of users is much higher than the number of developers.  It's impossible that all those users would agree with the designer's ideas about the best interface, but they have no chances to change it: they have to work with whatever they were given.

There is only one way to soften the conflict between the developers' decision of what is the best and the users' expectations: to give users some instrument of adjusting the applications to their tasks.  At the moment, there are two main directions in solving this problem: the *dynamic layout* and the *adaptive interface*.

Microsoft put huge efforts into the development of dynamic layout, and there are interesting results, but from my point of view the use of dynamic layout for the general design of applications is a big step in the wrong direction.  The dynamic layout is simply an automatic response to the problem of needed resizing, when the screen size or the screen font is changed.  The dynamic layout can be good on a local level, for example, inside a group, but it gives nothing, if the developer's and user's views of the form's design are simply different.  The dynamic layout enforces the designer's idea of the best elements positions and sizes regardless of any outside changes.  Users of such programs can't change anything.  It's like implementing a rule that, depending on the outside temperature, your shirt is predetermined from today and forever.

The adaptive interface is widely used for more than 20 years; there are a lot of papers, which propose different solutions for different purposes and parts of the interface, but there is one general idea behind all the forms of the adaptive interface. Somewhere inside, the designer makes the list of possible scenarios and decides about the best solution for each of them. Then users are given the list of choices, but on any selection the reaction was predetermined by the designer long ago. Programs become multivariant, but they are still absolutely ruled by developers.

The inner world of the programs is populated with the graphical objects and controls; elements of the first group are used for visualization of data and results; elements of the second group are used for transferring the users' commands to the programs.  Users maybe not even familiar with the term *controls*, but they know well enough how to select an item in the `ListView`, to put the mark in the `CheckBox`, to pick up one of the `RadioButtons` or to click an ordinary `Button`. The reaction on any of these commands was predetermined at the design stage and does not depend on the screen position of one or another control.  Then why all these elements are fixed?  If the button is used to start some process, then it absolutely doesn't matter, what size it has and where it is positioned.  Only the reaction on the button click is important for the work of an application.

Each user has his personal taste, personal style, and personal estimation of the good interface.  The important thing for interface design is not to be implemented according with the current day programming fashion, but to be the best from the point of view of each user.  Can a developer implement up to the tiny details an interface that would be the best for anyone in any circumstances?  Certainly not, because the number of users' preferences is infinitive and they are unpredictable.  But if any complicated interface can be rearranged by each user at any moment in a very simple way and in an instant, then it would be the best possible interface for everyone.  What has to be done to receive such a result?  Any screen element must become moveable and resizable.

Sounds easy and obvious, but up till now there were no such programs.  I want to show, how the programs are changed, when they are based on the moveable / resizable elements.  It's an absolutely different world, which can't be understood and estimated simply by reading an article.  Even if you are the best interface designer in the world, you would not understand the difference that such programs bring to the users until you'll start such an application and look at it in parallel with the reading.  All the samples of this article are from the **Test_MoveGraphLibrary** application.  At [5] you can download the ZIP file with such name that contains the whole project, ready to be used in Visual Studio; all the source files are written in C#.  For the purpose of pure demonstration, it's enough to take from this file the **Test_MoveGraphlibrary.exe** application together with the **MoveGraphLibrary.dll**.

Articles about the new ideas are usually written in such a way: idea – algorithm – some results – future development and expectations.  I want to write this article in an opposite way, so let's begin with the results and at the end look at the basis of this implementation.

Let's begin with the well-known Calculator program (**figure 1**) that sits on every computer with any version of the Windows system, so this program is used around the world by millions of people. I don't know your personal feelings about the design of this program; maybe you are quite happy with it. But somehow I think that you simply got used to it and never even thought about the possibility of changing it.

With me it is different. I use this program very rarely (once in two or three months), when I have to divide two big numbers. On those rare occasions I can find the buttons with the needed digits after some time, but the location of these buttons looks a bit strange to me. I also never use memory operations, so there are redundant buttons, which are more obvious (because of the red color) and attract my attention, keeping it from the digits that I need to find. I would like to get rid of those redundant buttons, but I can't, because the design is fixed. But maybe I can do something.

In the **Test_MoveGraphlibrary.exe** application, there is a similar Calculator (menu position *Nodes and covers – Calculator*) with the default view absolutely equal to standard Calculator (**figure 1**). The only addition, which is not demonstrated in any visual way, is that all these controls are moveable, so in a minute I rearranged the view to what I really need (**figure 2**). I simply change the size of the form and move any button by grabbing it with a mouse at any point of the button's border. The buttons that I don't need are moved out of the view across the form's border. This is not one of the predetermined views of my Calculator; there is simply no such thing as predefined views here. The view is automatically saved on closing the application; the next time it will be the same, as I left it. If I want to move the controls anywhere else, I can do it at any moment. Those relocations do not affect the work of the Calculator, as the positions of all these controls do not affect the code, linked with any click.

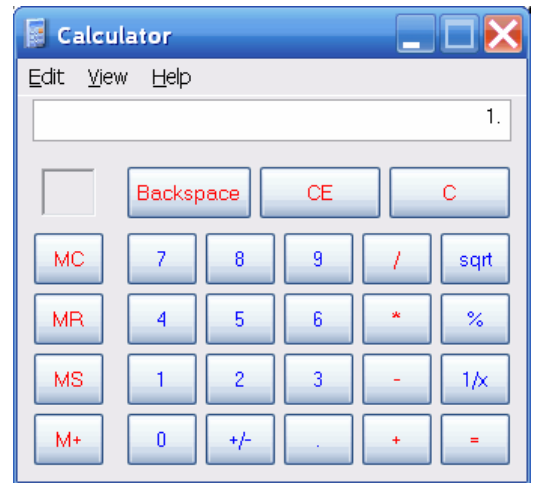

**Fig.1**  Calculator program

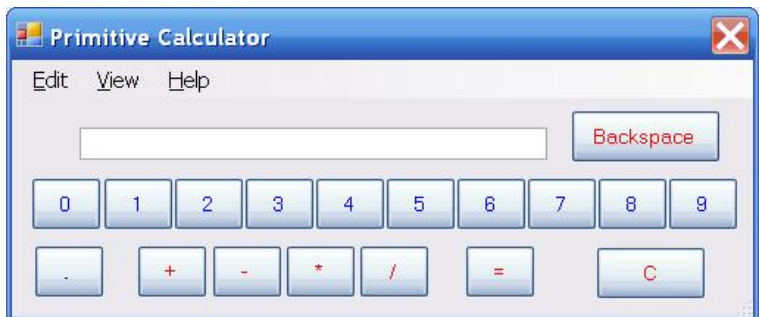

**Fig.2**  My preferable view of the Calculator

And here is the code, which turned the standard Calculator into fully ruled by users.

```
Mover mover;
mover = new Mover (this);
foreach (Control control in Controls)
{
    mover .Add (control);
}

private void OnMouseDown (object sender, MouseEventArgs e)
{
    mover .Catch (e .Location, e .Button);
}
private void OnMouseUp (object sender, MouseEventArgs e)
{
    mover .Release ();
}
private void OnMouseMove (object sender, MouseEventArgs e)
{
    if (mover .Move (e .Location))
    {
        Invalidate ();
    }
}
```

This is ALL! You can try this code in any of your programs in any form, populated with the controls; everything will become moveable. (Only don't forget to use and reference the mentioned DLL.)

This shows not only the implementation of the algorithm with the new programs, but how the already working applications can be changed. There are a lot of the widely known and well working applications, which do not require any changes in their working performance. These applications can be greatly improved by such an addition; the sample of Calculator demonstrates it.

The next figure demonstrates not the widely used program, but the well-known situation (**figure 3**). This is a case, when you have some data, of which you need to select a part for further processing. Usually the full set of data is shown in one list; the selected items are included into another list. The code works perfectly; you can select, add, and exclude the items without problems. There are no questions about these operations; the problem is in design: half of the users want to see the full list of items on the left and the selected items on the right; another half of users prefers those lists to be positioned in the opposite way. There is also some percent of those, who would like to see both lists in one column; they have the right to demand such positioning.

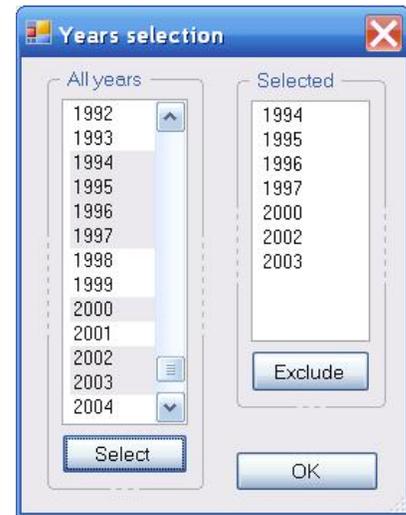

**Fig.3**   Selection can be organized
from any place to any

As a developer of such program, you can either position those groups in the way you prefer without any explanations, or you can try to explain to the users, why your vision of this simple interface is better than any other. It doesn't matter, what you prefer and what you fixed in the code: there are always those users that have different opinion. So why not to design this form in such a way that EVERYONE will position the lists in the way he personally prefers? What changes in the code are needed in order to add such flexibility into your application? The groups must become moveable / resizable and the OK button must be moveable.

Turning a button into moveable is already demonstrated in the case of Calculator; the same can be done with any `GroupBox`. By declaring the `MinimumSize` and `MaximumSize` properties of such group, you will make it resizable; anchoring of the group's inner elements will make it looks good at any size. This is an easy solution, but I don't like it. The problem with an ordinary `GroupBox` is that it can be moved only by the border, but not by the inner points. To make the situation even worse, the invisible border is close enough to the visible frame on three sides, so on those sides the `GroupBox` can be grabbed for resizing "by a frame", but on the upper side there is a significant gap between the visible frame and the invisible border. You can't grab such group "by the frame" there, but only somewhere far above (the change of the mouse cursor signals the possibility of catching an element). Because of these annoying things, I never use `GroupBox` in design, but prefer to use the similar `Group` class. Objects of this class can be moved by any inner point and resized by any frame point; the dashed parts of the frame inform about the possible type of resizing. **Figure 3** shows the **Form_DataSelection.cs** (menu position *Nodes and covers – Data selection*), where you can find all the needed code.

The OK button in this form is also moveable, so this helps to solve another, maybe a minor problem, but very annoying to some people. To close a form with simultaneous saving of some needed results, you often have to click some button (usually it is OK, Yes or Save). Developers position this button according with their own preferences; the most common places are in the middle at the bottom of a form, in the right bottom corner or in the right top corner. Users often think that the best place for such a button would be not where the developer positioned it, but somewhere else. In this form with a moveable OK button, any user can position this button in the best place personally for him. On closing the form, the button's position is saved, so this is done once and does not annoy you any more, but… I found out that with all the elements in the programs turned into moveable / resizable, you often want to move the buttons and other controls. Maybe a bit, but the best placement of the controls looks different on any new day. It's like moving the chairs at home: we always put them at nearly the same place, but not exactly at the same spot.

All the elements in the shown form are moveable / resizable. What is the code behind such a thing? To avoid repeating the same things again and again, I would like to underline two things.

1.  The object of the `Mover` class supervises the whole moving / resizing process; declaration and initialization of such an object is usually done in the same way, as was shown in the previous sample. A single mover is enough regardless of the number and types of moveable / resizable objects. It's a very rare situation, when the use of more than one `Mover` in the form is preferable; special situations are described in details in [4].

2.  Only three mouse events – `MouseDown`, `MouseMove`, and `MouseUp` – are used to organize the moving / resizing of all the objects. The left button is used to start the forward moving and resizing; the right button starts the rotation. In a lot of cases the code for these three events is exactly the same, as was shown in the Calculator sample.

So, is there any difference in making everything moveable in the **Form_DataSelection.cs** in comparison with Calculator? Yes, because not all the elements here are the controls. All the standard controls have a rectangular shape; the clicks and double clicks in their inner area are strictly linked with some reactions; users know these reactions very well and expect that the same controls in all the applications react to the mouse events in absolutely the same way. The design of all the applications is based on these strictly enforced rules, so the inner area of the controls is forbidden for anything new. Thus the only way to move and resize the controls is to use the frame around their borders.

The situation with the graphical objects is different. First, there are no strict rules for their reaction to any mouse events. Second, the variety of graphical objects is infinitive, so they can't be turned into moveable / resizable automatically. Something must be done with them (some code must be added). I'll write about this addition further on, and the whole design of moveable / resizable objects is described in details in [4], but there are two main ideas in design of such objects:

- Any object can be moved by any inner point.

- Any resizable object can be resized by any border point.

At the moment let's decide that all our graphical objects are moveable / resizable (for the demonstrated application it is really true!) and see, how it is used and how such change influences the design.

In the **Form_DataSelection.cs**, shown at **figure 3**, there are two `Group` objects; their initial sizes and ranges of resizing are defined through the parameters at the moment of initialization.

```
groupAll = new Group (this, rc, new RectRange (sizeTitles [0] .Width + 20, 180,
                                100, 500), titles [0], MoveAll);
mover .Add (groupAll);
```

Thus organized group is registered with the `Mover`, and everything else is exactly the same, as with Calculator.

Two samples, shown at **figures 1** and **3**, are fairly simple, but they demonstrate the principle difference between the standard our day applications and the new type of user-driven applications, based on moveable / resizable elements.

The well known Calculator from Microsoft has two different variants of view: standard and scientific; users can select only one or another. This is a classical example of an adaptive interface, when users have a list of choices, but the view for each selected case is absolutely determined by the designer. If I would organize the year selection on the ideas of adaptive interface, it would be easy to provide the selection between the shown view (**figure 3**), the mirror view, and something else. *User-driven* applications are principally different: there is no set of choices, but there is an instrument for the users to organize easily any needed view. As a designer, I don't determine the positions and sizes of the elements. I only guarantee that the selection will go correctly regardless of the positions and sizes of all the elements.

Let's turn to more complicated case and see, how such technique works to satisfy the users with absolutely different and even opposite demands

A lot of programs work with some set of personal data; sometimes the users have to enter their information through different controls; in other cases users receive the information from the database and look at the needed data in those fields. The exact set of controls depends on the purpose of application; the view of the form depends on the designer. The default view of the **Form_PersonalInfo.cs** (menu position *Nodes and covers – Personal information*) doesn't pretend for being better than many others; it simply shows the set of `TextBox` controls, which is common for such type of forms (**figure 4**). Each `TextBox` here is combined with some short text, which explains the particular piece of data, shown in this control.

**Fig.4** The default view of the **Form_PersonalInfo.cs**

The demonstrated placement of the controls is good enough for some western countries, but definitely not good for other parts of the world, in which the same data is expected to be shown in the different order. There are countries, in which:

- The family name always goes first.

- The day always precedes the month in any form of the date.

- The address is written in the opposite order: ZIP code, country, town, and so on.

Certainly, you can enter the needed information according with this form, even if it is very unusual for you, but what if you have to use it again and again? What if somebody has to analyse a lot of data from the database and this order of information is opposite to natural (native) way, to which he got used for many years? It becomes very annoying and the effectiveness of such work is very low. It is also possible that some users need only part of the whole information, so it would be nice to get rid of the unneeded information and not to show it at all; it would be nice simply to move those unneeded controls out of view.

There can be a lot of different cases; the designer can't predict them all and give the best predetermined solution to every user. It is much easier and much better to turn all these controls into moveable (and a lot of them also resizable) and let each user quickly and easily reorganize the form to the needed and best view. Each pair "control + text" in this form is an object of the `CommentedControl` class. Such object can be moved around the screen; the text can be also moved (and rotated) individually, thus changing the relative position of the control and its explanation.

All these `CommentedControl` objects are moveable; a lot of them are resizable. In the sample of my Calculator, the form's view can be changed and saved / restored, or the default view can be reinstated. Here you have much more. Any number of different views can be organized and saved for later use. The designer is not supposed to decide, how this form must be used, how many different views you are going to organize, and which part of the information is going to appear in any view. The designer only provides an easy way to save / restore the views, the easy to use instrument of moving / resizing all the parts, and guarantees that each piece of data will be shown in the appropriate field. The users decide about the part of information they want to see and the placement of each of those pieces.

If the form includes a lot of different elements, then the rearranging of such form can raise a question of how long it can take. It's nice to have an easy instrument for changing the view of each `CommentedControl`, but, for example, the address at **figure 4** is represented by the five objects of this class; if you want to move all these objects to another place, it would be nice to do it in one movement, but not in five.

**Figure 5** shows the **Form_PersonalInfo_DependentFrame.cs** (menu position *Nodes and covers – Personal information(III)*) and demonstrates one of the possible solutions for such a task. Each `TextBox` with some short explanation is of the same `CommentedControl` class, so there is no difference in moving and rearranging each control. But the groups, used in this form, are of the `DependentFrame` class; this is a very interesting class for forms' design.

In case of the year selection (**figure 3**) the `Group` objects were used. The common feature for the `Group` class and several other classes for forms' design, included into the **MoveGraphLibrary.dll** and described in [4], is the use of the dynamic layout for positioning the elements inside the groups. Such groups are resized by changing the frame; the sizes and positions of the inner elements are determined according with the frame's size. It gives to the users the control over the placement and size of the groups, but the related positioning of the inner elements is determined by the developer.

**Fig.5**     Personal information on the `CommentedControl` and `DependentFrame` objects

The `DependentFrame` class excludes the developer's decisions from this process. All the inner elements are moved and rearranged by users; the frame is automatically calculated on the combined position of all the inner elements and simply represents the area, by which the whole group can be moved. This maybe the best type of the groups' design, but there can be different views; the designer has a lot of choices.

**Figures 1 – 5** demonstrate several samples of using the moveable / resizable elements to rearrange the form's view. The change of the applications according with the users' demands is called customization. Tons of papers on the subject reflect the importance of this problem. The ideas of organizing better interface, proposed throughout the years, remind me the different types of transportation systems in the urban areas.

- When all the bus routes are declared once and for all by somebody in the city council, it is an analogue of the <u>fixed interface</u>.

- When the surveys of population are organized, and somebody can take their results into consideration, but the changes are still made according with the views of this person of what is really good, then it is an <u>adaptive interface</u>.

- When the routes are automatically lengthened with the development of new suburbs or the traffic intervals are automatically changed according with the population density, then it is a <u>dynamic layout</u>.

All these transportation systems are centralized; all such applications are absolutely *designer-driven*. Is there an alternative?

- The decentralized system with everyone driving his own vehicle is the analogue of the applications on moveable / resizable elements. These are the <u>user-driven applications</u>.

The biggest question is: in what areas and for what purposes such applications can be used?

The design of different forms is one of such areas; **figures 1 - 5** demonstrate only several samples. Of those samples, the Calculator represents the real and widely used application; others were developed only for demonstration purposes. But further on I write about the engineering and financial applications, which require different and very complicated forms for tuning of their parameters. All those tuning forms are totally based on moveable / resizable objects.

Design of different forms requires the use of absolutely different basic elements. I have included into the **MoveGraphlibrary.dll** several classes that can be used for forms' design and cover the needs of the frequently occurring situations. This set of classes is constantly in development; the new more powerful classes are added. But you are not restricted only to the classes that are currently in DLL; the same DLL gives you an instrument to design any class of moveable / resizable elements you need. At the moment I use such a set of primitives for forms' design.

- Any control can be turned into moveable / resizable. The ranges for resizing are declared via the standard control's `MinimumSize` and `MaximumSize` properties. The class of such moveable / resizable controls is called the `FramedControl`.

- The widely used case of a pair "control + text" is represented by the `CommentedControl` class. The most common case of such pair is the `TextBox` with some comment at the side, explaining, which parameter must be entered in this field. Such text (explanation) can be organized in the form of a `Label` control. But any control can be moved only by its border, so instead of it the `CommentedControl` class uses the painted text, which can be moved by any point. The control in the `CommentedControl` can be resizable, if its `MinimumSize` and `MaximumSize` properties get the appropriate values. The textual comment to the control retains the same relative position to the control when the pair is moved or when the control is resized. The text can be also moved and rotated individually.

- The `CommentedControl_LTP` class implements another version of the "control + text" pair. It uses a **l**imited **t**ext **p**ositioning (_LTP), but maybe easier in its moving around the screen.

- The `LinkedRectangles` class organizes the synchronous movement of any number of arbitrary positioned rectangles. Some of these rectangles can be occupied by controls; others can be the graphical objects, in which some text or any other information can be painted. Objects of this class are nonresizable, but they are very easy to organize.

- There are different classes for groups' organization; the previously demonstrated `Group` class is only one of them. Objects of this class are moveable by any inner point. There are four different types of resizing, which is determined by declared parameters at the moment of initialization.

- The `DependentFrame` class is a group that is determined "from inside". The inner elements of such group can be moved and rearranged individually. The frame is always painted around the combined area of all the included elements and represents the area, which can be moved by any inner point.

With the use of these classes (and others can be easily developed, if they are needed) any form can be turned into user-driven, where only users decide about the view and fulfillment of the forms. For such applications, developers do not give users the selection between several predetermined choices, but give an instrument, with which users can reorganize the view in any way they want.

The described list of elements, used for forms' design, is not full; there are some other classes, which were not mentioned, because they represent the minor variations or I stopped using them, because the new classes look better to me. But they still exist in the MoveGraphLibrary.dll and represent some interesting ideas in forms' design. When any application is designed, you try to develop it in a single style, so the different and concurrent classes for forms' construction are not used together. But for the best understanding of those classes and comparison of different approaches, I developed the **Form_PanelsAndGroups.cs** and filled it with objects of the different classes that can be used as "blocks" for forms' design (**figure 6**). All of them are moveable, but in different ways; the majority of them are resizable. This is the best form to look at their behaviour, compare, and decide for yourself, which of them are the most interesting from your point of view.

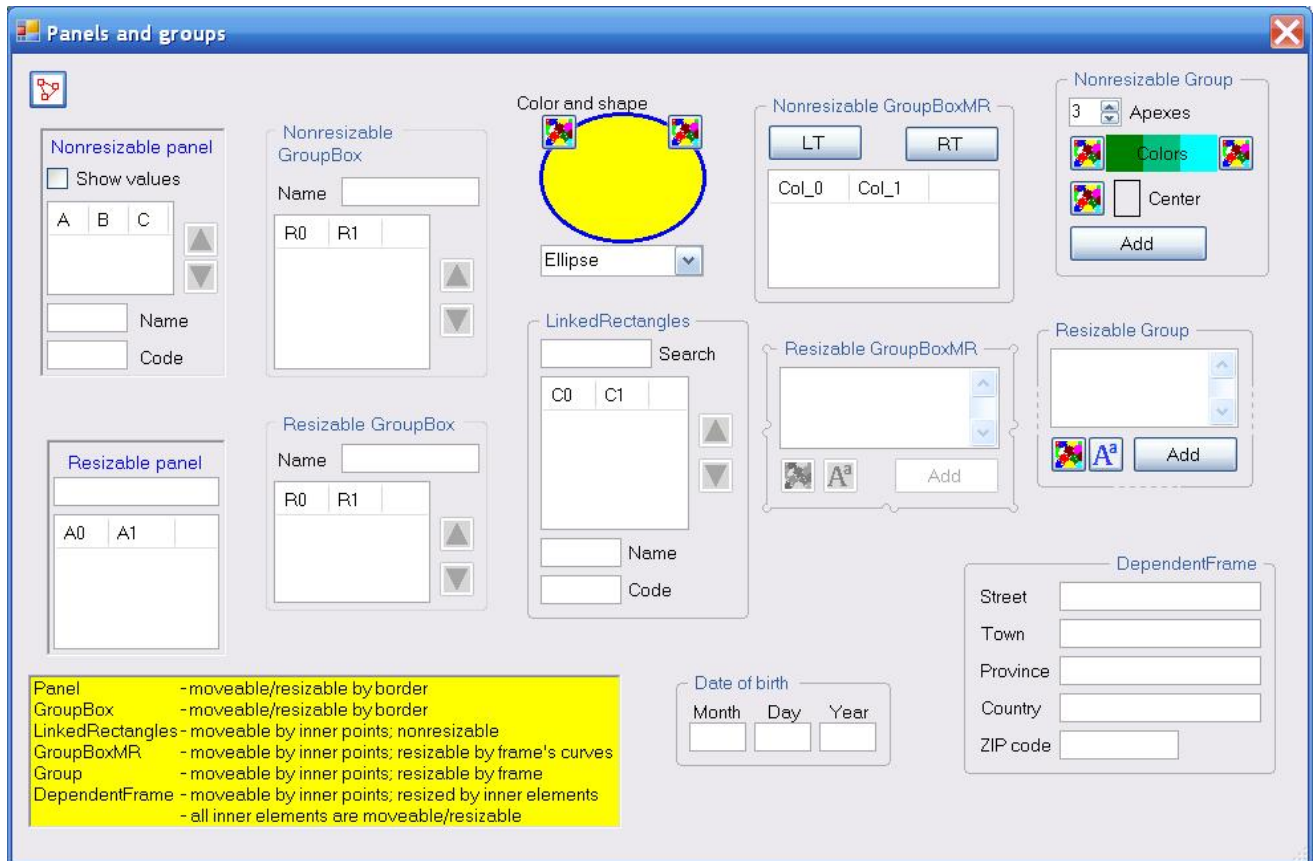

Fig.6 Form_PanelsAndGroups.cs

The implementation of new technique always puts a question about its usability in different situations. Even if it is easy to use in a simple case, is it the same for the complicated classes, or the complexity of the algorithm becomes inappropriate? There is no such problem with this algorithm; the turning of the simple rectangle into moveable / resizable is at the same level of complexity as for the very sophisticated scientific plots.

Some of the most interesting results are achieved in design of extremely complicated scientific and engineering programs on the basis of moveable / resizable elements. At the beginning, it was the area that triggered my work on the moveable / resizable graphics. In [1,2] I have already described a good working version of the moveable / resizable scientific plotting, and though it was only a year ago, the current version is now far away from that "old" one and the whole way of evolution perfectly demonstrates the rules, which come with the implementation of moveable graphics. Though it's a bit of a history, which is not always interesting to programmers, but in this case it demonstrates the basic rules of the applications on moveable / resizable elements, so its worth to look at the process of evolution.

The big scientific and engineering applications are among the most complicated programs, but their current day versions are nearly the same as they were 15 years ago; the strips and buttons can change from time to time according with the programming fashion, but the applications are mainly the same. Why is there no real progress in the area of such applications?

The design of such programs requires a lot of time and efforts; even the implementation of the users' requests often appears in the new versions only one, two or three years after the discussions. Even worse is the fact that the users of such applications are often much better specialists in the specific area than the developers. But all such programs are now designer-driven, so they are developed according with the understanding of the problems by lesser specialists and later used by the better specialists, who have no chances to go beyond that enforced level. Several years ago I came to the understanding of the fact that for such programs (and for many others) the idea of designer-driven applications is the impassable hurdle, which can't be crossed by any known or new form of adaptive interface. The only way to cross this barrier, to move ahead, and get more effective programs for engineers and scientists is the use of a different idea for applications' design.

The ruling has to be passed to users. This means that the main scientific (engineering) idea of any application is not changed and the inner mathematical instrument is still provided by the developer, but the control of the whole work from the basic things to the tiny details can be passed to the users. To organize such a thing, all the parts must be turned into moveable / resizable. I began to work on the moveable graphics, keeping in mind that the area of scientific applications must be the first to try the results.

When the first version of the algorithm for moving the graphical objects was designed, I immediately applied it to the scientific plotting. I took the big application with a lot of different plots, turned the plots into moveable / resizable, and began to play with this new program. It was a huge improvement of an application, but the more I worked with it, the stronger became the filling that something was incorrect. The logic of design was perfect for a fixed designer-driven application; even turning of a single part of it into moveable / resizable began to corrode the whole construction. The features, which were thought up to the tiny details and were used for years without any problems, didn't want to work smoothly with the new plots. The logic of moveable and non-moveable parts began to conflict.

The complicated plotting area is combined of different related parts; on the first stage only the main plotting area became moveable and resizable; the relocation of all the related parts were synchronized with the changes of the main part. Plots have different scales; usually the scales are positioned near the plotting area; in the old versions users could open the tuning forms and type in some parameters, which would change the scales' positions. The natural (with a mouse) moving / resizing of the plotting area changes the positions and sizes of the related scales in an appropriate way, but the individual relocation of the scales was still done only via the tuning forms, and it looked very awkward. If a plot with all its scales can be easily moved around, why the scales themselves can't be moved in the same simple way?

Certainly, I designed this next change, but it wasn't simply turning another graphical object into moveable. Plots and scales are the strongly linked objects with the type of "parent - children" relation. So now there were objects (plots), which could be moved synchronously with all their related parts, but at the same time those parts could be involved in the individual movements. That required some type of identification for the objects, involved in moving / resizing. This identification had to guarantee the correctness of synchronous and individual movements for the objects with any type of relations between the parts. It had to be not the new type of identification for each new class of objects, but the general solution for any classes, which will be designed in the future and involved in different types of movements.

With the plotting areas and scales now moveable in any possible way, my attention turned to another huge problem, for which neither I nor anyone else could find a good solution before: positioning of the comments. In all the numerous scientific and engineering applications the plotting areas were better or worse positioned by the designers in the old way either by declaring their coordinates, or via implementing the dynamic layout, or with some other type of adaptive interface. The results of calculations, which are shown in those areas, often need some comments, but the exact lines' positions are calculated throughout the work of applications, so there is no way to determine beforehand the good position for these comments. Again, positioning of the comments was done before via declaring them in some tuning forms, but now it looked very strange when the plots and scales could be moved easily around the screen, but the comments to these objects were positioned and changed in some archaic way. The comments have to be moved exactly in the same easy way, as plotting areas and scales – by a mouse. In addition, the comments need to be not only moved forward but also rotated for better placement along the arbitrary lines.

I think that now it is obvious in which direction the scientific applications started to change after a single element – the main plotting areas – was turned into moveable / resizable. There is a law in engineering that the reliability of the whole system can't be higher than this characteristic for the lesser reliable part. The similar rule, translated into the world of programming, declares that the flexibility of the system can't be higher than this characteristic for any part. If you have a single moveable element in the form, it will begin to demonstrate all the time the lack of such feature in all other surrounding elements and demand their change in similar way.

Even if you turn all the graphical objects (plotting areas, scales, and comments) into moveable and resizable, but leave anything unmovable, then the users would always bump on this hillock, even on a small one. The inner area of applications is populated both with the graphical objects and controls, so controls and groups of controls must be turned into moveable / resizable. I made this change, received the new type of scientific application (**figure 7**), and that was the new paradigm – ***user-driven application***.

**Figure 7** shows the **Form_ManyPlots.cs** (menu position *Y(x) plots – Many plots*); it is a relatively simple demonstration form, which is kept simple for better understanding. But at the same time it demonstrates the features and design ideas, which are common for all the complicated scientific applications, which I design now.

The form's area is populated with four different types of elements.

- <u>Plotting areas</u> are moved by any inner point and resized by any border point. Areas can have any number of scales and comments.

- <u>Scales</u> can be moved individually by any inner point and positioned anywhere relatively to their "parent" plotting areas. Resizing of scales is done automatically according with their "parent" areas resizing. Scales can have any number of comments.

- <u>Comments</u> can be moved and rotated individually by any inner point. When the plotting area or scale is moved or resized, all its associated comments move in such a way, as to keep the same relative position.

- <u>Controls</u> are moved and resized (if needed) by their border points. The nature of

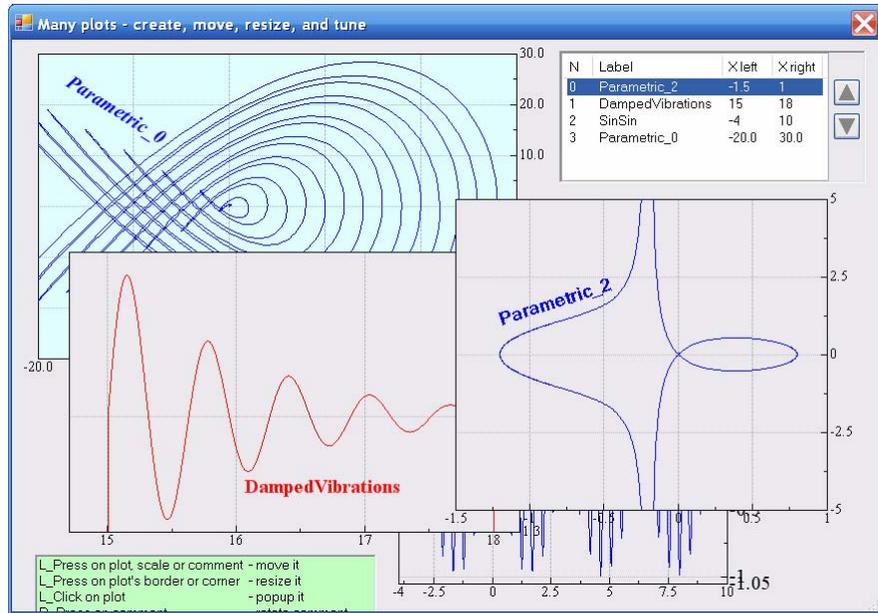

**Fig.7** Typical form of a scientific application, populated with plotting areas, scales, comments, and controls.

controls would not allow them to be moved by the inner points, because users know very well, what must be the standard reaction on pressing controls in any inner point, and this reaction must be not changed.

After all the inner objects of an application became moveable / resizable, the next step had to be done. I wasn't thinking about such a possibility from the beginning; one by one I was turning all the objects into moveable, but when it happened, I saw that the main design idea had to be changed. It wasn't planned ahead; it was the logical result of using the elements with absolutely new features.

- All the tuning or additional forms, associated with those moveable / resizable plots, must be also redesigned in the similar way, so that inside them everything also becomes moveable.

The big scientific and engineering applications require a lot of time and huge efforts for their design. If such applications are designed not for the inner use, but for the market, then the number of users is many times bigger than the number of developers. There are no chances that all the users would be satisfied with the proposed design, and the process of development is often accompanied the discussions, hot discussions or even quarrels between developers and users. Not a rare situation, when users have opposite interface requirements; this is a dead end for any fixed interface, but for the new type of applications it's not a problem at all. Moveable / resizable objects simply eliminate this whole problem; there is nothing to argue about any more; any user can rearrange the view of an application in any way he wants. The problem of annoying interface must be closed.

This is an obvious and not the most important feature of the user-driven applications. More important is that they are turned into the instruments of solving problems in one or another area. Our current day applications, even the most sophisticated, are in reality only the very powerful calculators. Turning these applications into the instruments, tuned by the users, give those users a very flexible research instrument. Users of the scientific and engineering programs are often much better specialists in the specific area than the developers of the programs, they are working with. The work of the better specialist must not be limited by the level of the lesser specialists.

**Figure 7** is from a demonstration program, which allows to analyse any function. As a developer, I don't know how each function would be calculated (there are five different ways of receiving numbers), what screen they are going to use, how many plotting areas the users would like to organize, how these areas would be positioned, how many different functions would be shown in each area, and what would be the viewing parameters of each area and each piece of data. I designed an instrument of analysis, which means:

- Arbitrary form sizes.
- Any number of plotting areas in the form.
- An arbitrary number of functions in each plotting area.
- An unrestricted moving and resizing of the areas.

- An easy way of changing all the visualization parameters.

- An easy way of adding, deleting, modifying, and moving the comments.

- An easy way and many variations of saving and storing any piece of visualized information.

There is not a single restriction from the designer's side. It's an absolutely user-driven application. Do whatever you need and want to. Could you think about such an application before? No, because there was no basis, on which it could be constructed. And the only required thing to start the design of such programs was an easy to use instrument of turning any object into moveable / resizable.

One more sample from the area of scientific applications – the form from the DataRefinement application (**figure 8**). A lot of researchers receive the needed data from the analog-digital converters. This data is going to be analysed by a well known mathematical methods, but the data is often obtained together with an additional noise, and the well known methods of mathematical analysis go crazy, while trying to solve the equations with such input values. The data has to be previously refined by a researcher, and that's where the user-driven application can be used at its best. The refinement of such data means that some subsets of the data must be substituted by very close values, but obtained as a result of very simple calculations: straight line, parabola, etc. The selection of segments and the type of the simplification can be done only by the researchers throughout the visual analysis of the data.

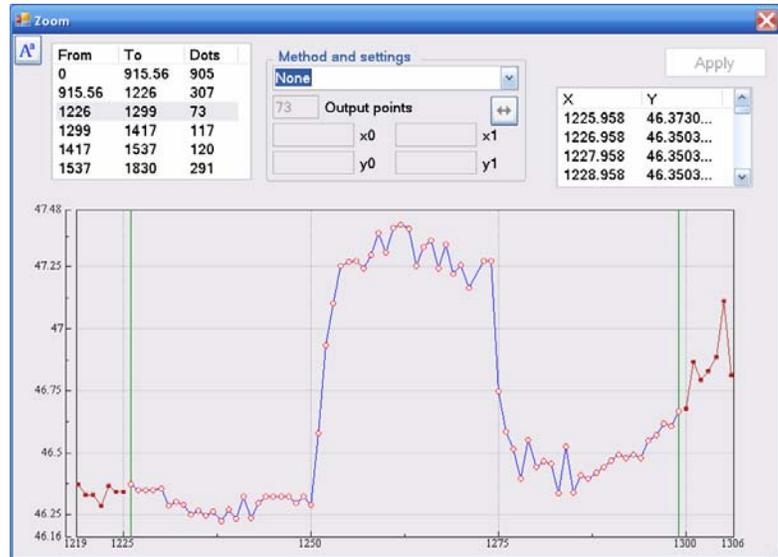

**Fig.8** One form from the Data_Refinement application.

It can be done only manually and is based on the experience. It is the classical situation, where the results of mathematical analysis must be used in parallel with the manual adjustment.

The specialists, who commissioned me to develop this program, thought about it for years, but never designed it. The whole task demands the manual manipulation of data, so the good application of such type needs the different movements of the visualized data. It took me a very short time to develop several classes of differently moving objects and combine them with the ideas, which are general for all other scientific applications.

The only thing, which had to be discussed, is the protocol between the input data and the visualization; from there on the users can do whatever they want (need) with this data. The data can be segmented in any possible way by any number of moveable sliders. Each dot, representing the pair {x, y} of an input data, can be moved up or down in order to get rid of the noise. All the objects in the form are moveable / resizable, so the design ideas do not interfere with the analysis.

In the previous sentence I mentioned as an obvious thing that "all the objects in the form are moveable / resizable". It's obvious and absolutely natural to me, because all my programs work in such a way, but it is absolutely new to nearly every reader of this article. For decades I design the very complicated scientific applications, so for many years the process of design of such form (dialog) would go in a standard way.

1. Fulfillment of the form. This dialog has to include one plotting area, two `ListView` controls with the data, a group of several controls to declare the parameters, and a couple of additional buttons.

2. Positioning of the parts. The set of controls determines the size of the framed group; the amount of data to be shown in the lists determines their sizes. Depending on the relative sizes of these three elements, I would position them either in a row (as shown at **figure 8**), or in a column. In case of a column, the width of all three elements would be set equal; such column can be placed at the left or at the right side. The remaining part of the form will be occupied by the plotting area.

3. Adaptation. Anchoring will be used as a reaction to the form's resizing. In some proportions the whole design will look better than in others.

This is the standard process of forms' design for programmers around the world. Some people were born with the feeling of the good form and harmony, others were taught a bit in this area. There are good programmers and there are not so good; some are even worse than can be imagined. Users have no choices: they have to work with what they are given; they are lucky, if the programmer was a good specialist. If not, users will still have to work with this application, but I'll omit the words they would be pronouncing in the address of such a designer.

The difference between the numerous similar applications and the form, shown at **figure 8**, is that the lists, the group, the buttons, and the plotting area are moveable. Lists and plotting area are also resizable. As a developer, I will produce something, which is good from my point of view, but I do not insist that everyone looks at the world with my eyes. Users can change the form in any way they want; I am happy to design an instrument, which allows everyone to use my applications without any discomfort. Any changes you make will be stored until the next time you'll open the same form. If you prefer this form to look different on a sunny day than on a rainy one, it's absolutely OK. The main thing here is the data analysis. That's the main goal of this application; all the annoying things must be excluded.

The difference with the standard design is simple. From the beginning the form must be looked at as a collection of the moveable / resizable elements. The purpose of each object is not changed at all, the reaction on user's selection of any parameter is unchanged, so the main part of the underline code is the same. The additional steps for the new design are exactly the same, as were shown in the case of Calculator.

1. Declare the `Mover` object that will supervise the whole process.
2. Register all the moveable / resizable parts with this `Mover`.
3. Use the same three mouse events.
4. Add the code for saving and restoring the needed sizes and positions.

It looks like the scientific / engineering applications are just the programs to be turned into user-driven and in which the ideas of such design must be implemented everywhere from the tiny details to the level of the main design. But this is not the only area, which can benefit on these new ideas. Applications with financial graphics are not less interesting.

**Form_Medley.cs** (menu position *Financial graphics – Medley*) demonstrates the simultaneous use of different types of financial graphics (**figure 9**). It's again not the application, designed according with some strict requirements, but an instrument of data analysis, and from this point everything is in the users' hands.

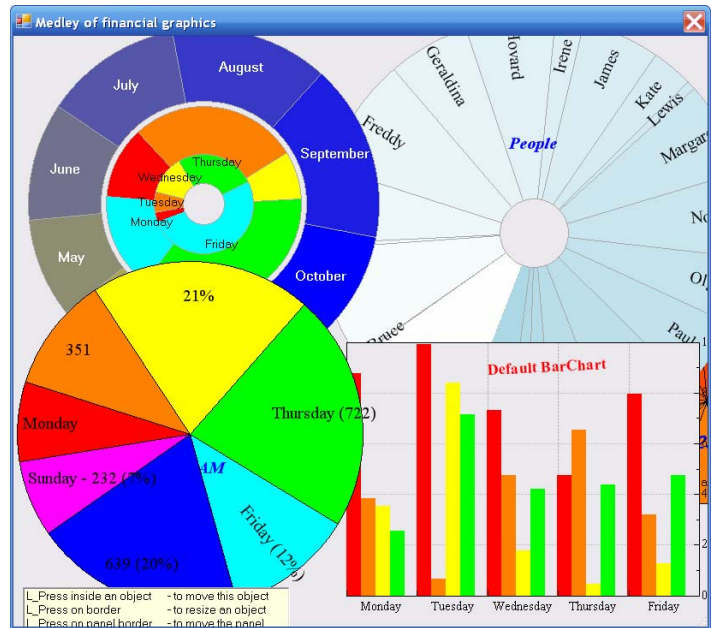

**Fig.9** Medley of financial graphics.

- The number of objects is determined only by the user's wish; objects can be added or deleted at any moment via a context menu.
- Any object is moved by grabbing it at any inner point.
- Any object is resized by any of its border points. For a set of the coaxial rings this is applied also to the borders of each ring.
- Each class of financial graphics has its system of textual information; some of the texts are associated with the whole object, others – with the parts of objects. Each piece of textual information can be moved and rotated individually; it is also involved in synchronous movement with its "parent", when the last one either moved or rotated.
- All the visualization parameters, and there is a lot of them, can be easily tuned. The tuning can be done on an individual basis, or the parameters can be spread from one, used as a sample, on all the siblings, or the parameters can be spread to all the "children".

The variety of the possible views is infinitive; the number of elements on the screen is unlimited. Such unrestricted flexible application with a huge variation of parameters can work only on the basis of absolutely reliable system of objects' identification. The identification of all the tiny parts and complicated objects is organized in the same simple way as was done for the classes, used in the area of scientific applications. There is a single `Mover` and the same code in the three mouse events. A simple code produces an astonishing result. User-driven application!

I have demonstrated the samples from different areas. The common thing is that all the elements of such applications are moveable / resizable; this moves such programs into special class. Now it's time to look into the algorithm that allows to do it.

The main idea of the used algorithm and its many additional features are based on the correct formulation of the problem, which I was solving.

- I need to move an object by grabbing it at any point, so the whole object must be covered by a set of sensitive areas, which I call *nodes* (class `CoverNode`).

- Objects can be of any possible shape, so the proposed set of nodes (class `Cover`) must allow to cover the arbitrary form.

- The resizable object must be grabbed for resizing at any border point, so the whole border must be also covered by a set of nodes. If by the idea of design the resizing of an object must be allowed only partly (for example, in one direction), then only the corresponding part of the border needs to be covered.

- With the arbitrary form of objects and the limited set of the node shapes, there will be samples, which would be impossible to cover with such nodes, if the nodes have to be placed only side by side. But the covering is much easier, when the nodes can overlap, so the overlapping of nodes is allowed and has no negative effect on the whole algorithm.

- Some objects may consist of the separate parts. These parts can be involved in the individual or synchronous movements, so there must be a mechanism to organize both of them and not to lose the links between the parts, when they are moved individually.

- Usually the screen objects would be involved in the forward movement, but some of them need to be rotated. Both movements can be started at any inner point, so they are distinguished only by the pressed button; I use the left button to start the forward movement and the right button for rotation.

- Addition of moveable / resizable features to all the screen objects must be done without any changes of the objects' view. The indication of possible movement is done via the changes of the mouse cursor, but usually this is all. Users KNOW that everything is moveable / resizable, and this knowledge is just enough. (There is no other indication of the possibility of windows' moving and resizing on the upper level; you simply know that it can be done and use it.)

- The indication of the moving / resizing possibility can be added to the screen view of the objects, but this is an additional thing, which can be decided by the users. The whole mechanism of moving / resizing works regardless of whether this additional visualization is ON or OFF.

There can be different solutions to implement such set of rules. Here are several remarks about my algorithm.

Any class of objects, which are turned into moveable / resizable, is derived from the `GraphicalObject` class that declares three crucial methods.

```
abstract class GraphicalObject
{
    abstract void DefineCover ();
    abstract void Move (int dx, int dy);
    abstract bool MoveNode (int i, int dx, int dy, Point ptM, MouseButtons mb);
```

The cover of an object consists of a set of nodes and is defined in the `DefineCover()` method. A node can be either a circle, a strip with two semicircles at the ends, or a convex polygon. Any node may have an arbitrary size. In addition to original location and size, each node has two characteristics: the type of possible individual movement of this node (Any, NS, WE or None) and the form of the mouse cursor, when it moves across this node. **Figure 10** demonstrates the cover for a chatoyant polygon, which can be involved in all possible types of transformation: it can be moved and rotated by any inner point, it can be reconfigured by any apex or center point, it can be zoomed by any border point.

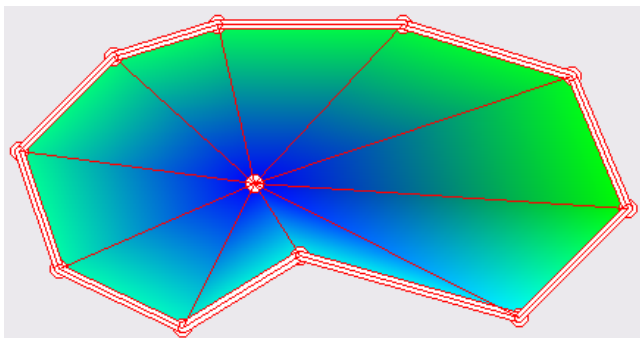

**Fig.10**   Cover of the chatoyant polygon demonstrates the use of all three types of nodes. Polygon can be moved and rotated by any inner point, reconfigured by any apex or center point, and zoomed by any border point.

The `Move()` method describes the forward movement of the whole object. The drawing of even the most complicated objects is usually based on a few primitives like points and rectangles; the `Move()` method describes only the movement of these primitives.

Usually even the covers of the most complicated objects consist of a small number of nodes, but for the curved borders the special type of *N-node* covers is used; such covers may include a huge number of small nodes with identical behaviour. **Figure 11** demonstrates this type of covers for a circle and a ring.

The `MoveNode()` method describes the individual movement of the nodes; it must be defined only for the nodes that are involved in such movement. Interestingly that the objects with the N-node covers, which can have a significant number of nodes, have very short `MoveNode()` method, because all those nodes behave in a similar way.

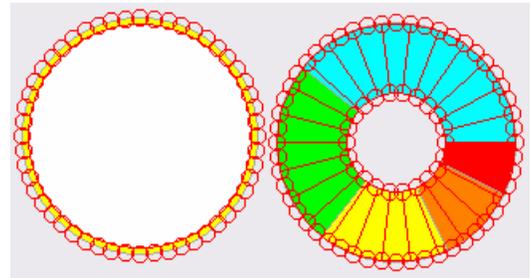

**Fig.11** N-node covers for a circle and a ring.

The special `Mover` class is developed to supervise the whole moving / resizing process. It doesn't matter how many moveable objects are in the form (dialog), to how many different classes they belong, and how different are their movements. It's enough to have one `Mover` object in the form and register with it (include into its List) all the objects to be moved and resized; after it this `Mover` will take care of all the operations. All the moving / resizing operations are organized with a mouse, so only three mouse events are used: `MouseDown`, `MouseMove`, and `MouseUp`.

The above mentioned rules of turning an object into moveable / resizable are applicable to the graphical objects; the situation with the controls is a bit different and in many cases simpler. Users know, what to expect from pressing the mouse at any inner point of any type of control, so this reaction must be not changed. Thus the inner area of all the controls is forbidden for using it as a starting point for moving / resizing, and the only available area is the frame around the controls' borders. This boundary zone must be divided between the areas for moving an object and the areas for its resizing; in case of nonresizable object its whole border can be used for moving.

The standard form of any control is a rectangle, so the sensitive area must be of the rectangular shape slightly out of its borders. With the standard form of the sensitive area, the nodes for resizing must be always placed in the same way, so that they can be easily found. The cover for the objects of the standard form with the standard positioning of the nodes can be organized automatically, so there is no need for the `DefineCover()` method in case of controls; the cover can be provided by the `Mover` automatically.

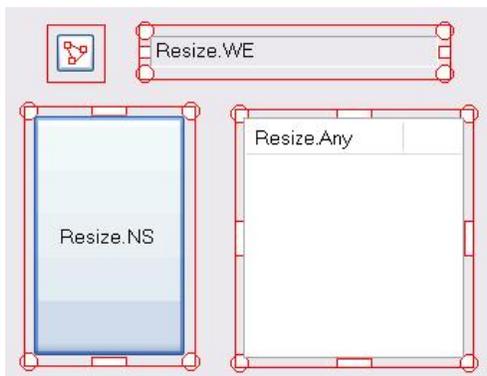

**Fig.12** Covers for the controls' resizing

**Figure 12** demonstrates the covers for all four different types of controls' resizing. The big red frame shows the area, by which the control can be moved; certainly, it is only the strips between the red line and the control's borders. The rectangular nodes in the middle of the sides are for resizing in those directions. Usually, covers are not visualized, so all these nodes inform about themselves only by changing the mouse cursor, when it is moved over them. Without covers' visualization, the finding of a relatively small node somewhere in the middle of a lengthy side can be a problem; to ease this search, the size of the nodes in the middle of the lengthy sides is increased. But there are always the nodes in all the corners of a resizable control; finding of these nodes is not a problem at all, so the resizing with these nodes is an easy thing.

The movement of a control means the simple change of this control's coordinates, so the `Move()` method is not needed at all.

The `MoveNode()` method is also not needed for the controls, because the numbers of the nodes for the standard covers are predetermined and the reaction on moving any node is predetermined also.

Thus all three methods, which are extremely important for the graphical objects, are not needed for the controls. As was shown in the first samples of this article, the only needed thing is to register a control with the `Mover`, and after it the whole moving / resizing is done automatically. The only needed thing for resizing is the declaration of the ranges for resizing via the `MinimumSize` and `MaximumSize` properties of such a control.

Thus briefly described technique is used to turn the screen objects into moveable and resizable; up till now I didn't find a single sample of an object, which would cause problems in turning it into moveable. You can read the most detailed description of this algorithm with a lot of samples in [4].

So what are the main changes on the way to user-driven applications, and what can be the consequences?

All the objects and groups of objects are turned into moveable / resizable. A proposed algorithm can be applied to objects of any form, behaviour or origin (graphical objects or controls). The algorithm itself is not so crucial, but there must be some way to do it easily and quickly. The best thing is to look at any object as moveable from the beginning of design and not as "maybe it will be turned into moveable later".

Stop thinking about any application as a finished construction that is going to stay unchanged from now on and forever. Look at it as a nice set in a furniture showroom: it must be shown in the best way to attract the customers, but after it is bought, the salesman is not telling where and how each piece must be used; it's the buyer, who makes all the decisions. You design an application; let users decide about the best way of using it.

These two things sound really simple and easy to understand. The problem is that it's not so easy to start thinking in such a way, because the step from the currently used applications to user-driven applications is bigger and more revolutionary than from DOS to Windows. 20 years ago a lot of people were saying, even without seeing Windows: "OK, what is this all about? Moveable and resizable windows? What for? We worked without it for many years; we can go on without it for years more." Certainly, some people can. But it's impossible to imagine a Windows system, in which not users, but the system itself would decide about the position and sizes of each window, which is going to be opened. That's the adaptive interface or dynamic layout at their best; that's how the applications are designed now. Why? Just because. I can't remember a single person, who would ask a simple question: "Why all these graphical objects are NOT moveable?" There are hundreds of papers about the different adaptation techniques, but there was not a single attempt to make **everything** moveable. Why? Tradition. Nobody declared that it was possible.

Well, now there are samples of such applications and an instrument to construct them, you can try and see how it looks and works. There are two things that happen, when you start developing the user-driven applications and working with them.

1. If you are a programmer and give your clients a single application of such type, there is no way back for you. After it, your clients will never accept any application, which they can't redesign in any possible way they want.

2. Soon you'll expect that all the objects in all other applications are also moveable and you will be very much surprised to find, if they are not. In all the applications you'll try to move the inner objects to the better positions. This is the best indication of the fact that we need such objects in all the programs.

To describe this new type of programs, I use all the time the term ***user-driven applications***. There can be another term – ***personal applications***, and I am not sure, which of them is more precise. The purpose of each application is not changed and it continues to work according with its main goal (problems of some scientific area, financial analysis, and so on), but any user can make such drastic changes with any application that even the designer would hardly recognize his product. Never mind; the application works, it solves the problems, for which it was designed, and it takes exactly the view, which user prefers. That is the main thing. Application is simply personalized.

Programming is the reflection of the real world on the screen. Very few of the things that we constantly use, are really unmoveable; the majority of the things can be moved, and we use this feature whenever we need. Exactly the same thing must be done to the objects inside our programs.

Dr. Sergey Andreyev ( andreyev_sergey@yahoo.com )

June 2009